\documentclass[lettersize,journal]{IEEEtran}
\IEEEoverridecommandlockouts
\pdfoutput=1

\usepackage{url}
\usepackage{cite}
\usepackage{makecell}
\usepackage{subfigure}
\usepackage{subfig}
\usepackage{amsmath,amssymb,amsfonts}
\usepackage{amsthm}
\usepackage{algorithmic}
\usepackage{algorithm}

\usepackage{graphicx}
\usepackage{fancyhdr} 
\usepackage{textcomp}
\usepackage{booktabs}
\usepackage{bm}
\usepackage{float}  
\usepackage{dblfloatfix} 
\usepackage{pifont}
\usepackage{comment}
\usepackage{cellspace}
\usepackage{wasysym,makecell}

\usepackage{multirow}
\usepackage[normalem]{ulem}
\useunder{\uline}{\ul}{}
\usepackage{xcolor}

\def\BibTeX{{\rm B\kern-.05em{\sc i\kern-.025em b}\kern-.08em
    T\kern-.1667em\lower.7ex\hbox{E}\kern-.125emX}}
\allowdisplaybreaks

\begin{document}

\title{
Snake Learning: A Communication- and Computation-Efficient Distributed Learning Framework for 6G
}
\author{Xiaoxue Yu, Xingfu Yi, Rongpeng Li, Fei Wang, Chenghui Peng, Zhifeng Zhao and Honggang Zhang 
\vspace{-.5cm}
\\

\thanks{This work was supported in part by the National Key Research and Development Program of China under Grant 2024YFE0200600, in part by the Zhejiang Key Research and Development Plan under Grant 2022C01093, in part by the Zhejiang Provincial Natural Science Foundation of China under Grant LR23F010005, in part by the National Key Laboratory of Wireless Communications Foundation under Grant 2023KP01601, and in part by the Big Data and Intelligent Computing Key Lab of CQUPT under Grant BDIC-2023-B-001.}
\thanks{© 2025 IEEE. Personal use of this material is permitted. Permission
from IEEE must be obtained for all other uses, in any current or future
media, including reprinting/republishing this material for advertising or
promotional purposes, creating new collective works, for resale or
redistribution to servers or lists, or reuse of any copyrighted
component of this work in other works.}}
\maketitle

\begin{abstract}
In the evolution towards 6G, integrating Artificial Intelligence (AI) with advanced network infrastructure emerges as a pivotal strategy for enhancing network intelligence and resource utilization. Existing distributed learning frameworks like Federated Learning and Split Learning often struggle with significant challenges in dynamic network environments including high synchronization demands, costly communication overhead, severe computing resource consumption, and data heterogeneity across network nodes. These obstacles hinder the applications of ubiquitous computing capabilities of 6G networks, especially in light of the trend of escalating model parameters and training data volumes. 
To address these challenges effectively, this paper introduces ``Snake Learning", a cost-effective distributed learning framework. 
Specifically, Snake Learning respects the heterogeneity of inter-node computing capability and local data distribution in 6G networks, and sequentially trains the designated part of model layers on individual nodes. 
This layer-by-layer serpentine update mechanism contributes to significantly reducing the requirements for storage, memory and communication during the model training phase, and demonstrates superior adaptability and efficiency for both classification and fine-tuning tasks across homogeneous and heterogeneous data distributions. 
\end{abstract}

\vspace{-0.4em}
\section{Introduction}\label{sec1}
\begin{figure}[t]
	\centering 
  \hspace*{-0.3cm} 
	\includegraphics[scale = 0.46]{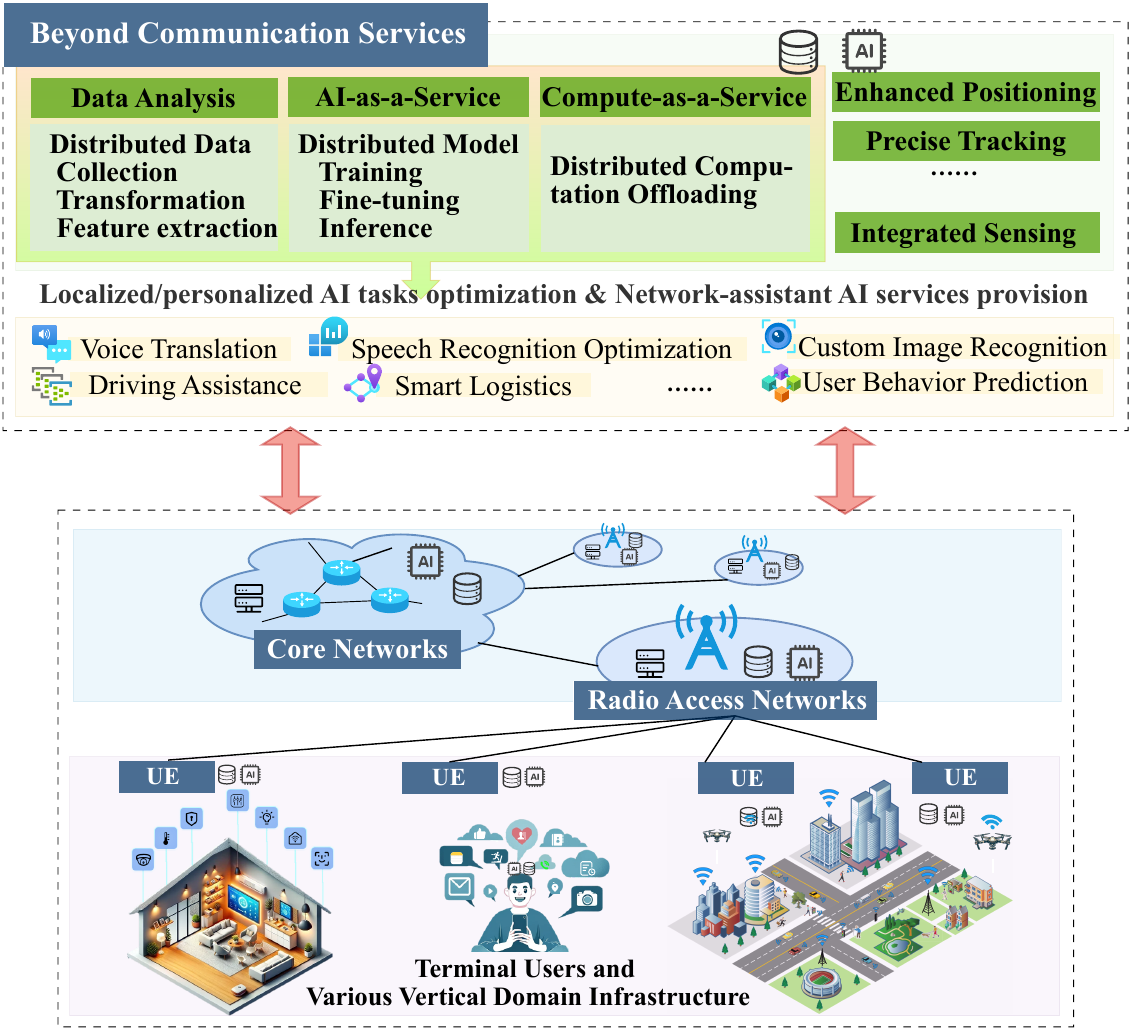}
	\caption{Examples of Beyond Communication Services Overview provided by 6G networks. }
	\label{fig:6GAI}
	\vspace{-1.em}
\end{figure}

As 5G networks lay the groundwork to support Artificial Intelligence (AI)-related operations, especially enabling Federated Learning (FL) and model distribution, etc \cite{3GPP}, 
6G is expected to further enable deeper integration of AI-related capabilities and communications, as shown in Fig. \ref{fig:6GAI}, by supporting both ``learning to communicate'' and ``communicating to learn'' paradigms \cite{Reviewer1SuggestPaper}. 
This advancement enables 6G to evolve into a dynamic, distributed computing platform that leverages intelligent User Equipment (UE) and Network Elements (NEs) as computing nodes, collectively offering Compute-as-a-Service (CaaS) and AI-as-a-Service (AIaaS). 
Such a transformation enhances distributed computation offloading, model training, and inference, further unleashing the potential of in-network resources while optimizing capital expenditure. 

Despite the apparent importance of an effective and efficient distributed training and/or fine-tuning framework, designing such a framework in 6G networks still encounters considerable unique challenges \cite{campolo2023network}. 
First, the inherent variability of the wireless network environment and the heterogeneous \&  dynamic computing resource availability 
can destabilize existing distributed learning frameworks (e.g., FL \cite{fedavg} or Split Learning \cite{sl}) and 
exacerbate the communication burden due to {their} frequent, real-time synchronization for parameter aggregation or intermediate activation and gradient transmission \cite{woisetschlager2023federated}. 
Although asynchronous methods can relax synchronization requirements, they give rise to certain complications \cite{kuang2023federatedscope} such as model inconsistency. 
Second, the statistical heterogeneity of data across scattered nodes, characterized by diverse distributions and sizes, can hinder convergence and degrade model quality.
Third, model training (fine-tuning) and inference in a distributed manner exhibit significantly distinctive demands on memory bandwidth and computing resources \cite{miao2024flexllm}, particularly in 6G networks where resources are shared among communication, CaaS and AIaaS functions. 
Traditional dynamic switching operations \cite{niu_tango_2011}, based on traffic ``tidal effect" in telecommunications, may no longer suffice to fully harness the capabilities of widely deployed computing resources in 6G. 
Therefore, a qualified training (fine-tuning) framework that efficiently leverages dynamically available communication and computational resources while ensuring model consistency is essential to optimize resource utilization and complement the development of AIaaS in 6G networks \cite{NetGPT}. 

In a nutshell, this article proposes a novel communication- and computation-efficient distributed collaborative learning framework for 6G networks, termed ``Snake Learning". 
Inspired by the ``Snake" game, where the snake grows by progressively consuming items, Snake Learning incrementally improves model performance by serpentine layer-wise updates across different computing nodes with their local data. 
Significantly different from existing approaches and frameworks like parallel-oriented FL \cite{fedavg}, relay-based Split Learning \cite{sl}, and many variants \cite{RWS,thapa2022splitfed,han2021accelerating}, the serpentine, incremental parameter update and transmission relax real-time and frequent synchronization requirements while maintaining privacy friendliness. 
To ensure seamless integration, Snake Learning meaningfully calibrates service components and workflows in 6G networks, and novelly incorporates several modules like data processing and Knowledge Distillation (KD). Benefiting from these revolutionary efforts, Snake Learning yields remarkably improved computation and communication efficiency and reduces storage and memory demands, making it well-suited to support model training or fine-tuning on resource-constrained nodes in 6G networks.

\section{Existing Distributed Learning Frameworks and Key Issues within 6G Networks}
\subsection{Existing Distributed Learning Frameworks}

\subsubsection{Federated Learning} 
\begin{figure*}[t]
	\centering 
 \hspace*{-0.24cm}
	\includegraphics[scale = 0.485]{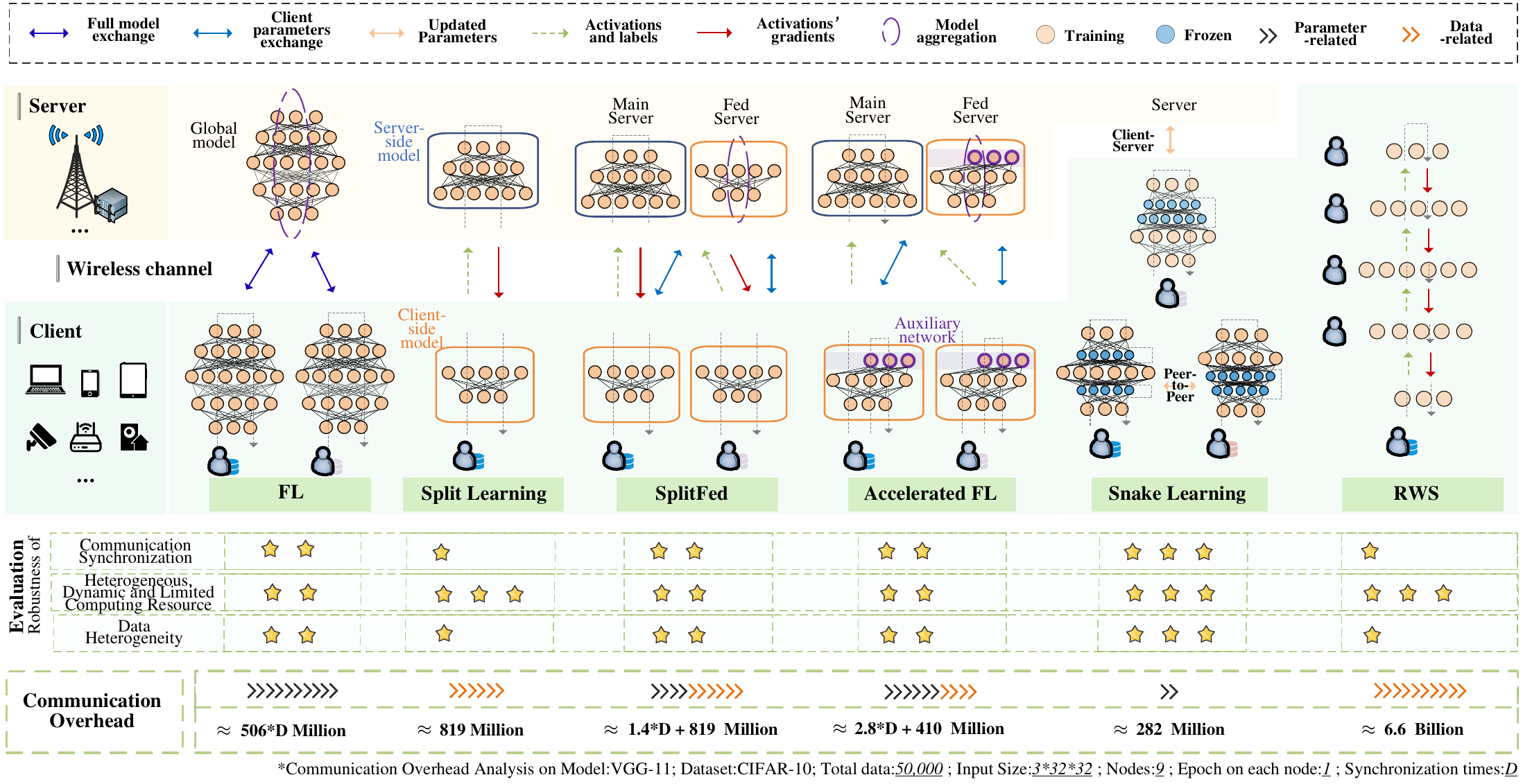}
	\caption{Comparison of different distributed learning frameworks.}
	\label{fig:ArchitectureAndComparison}
	\vspace{-1.2em}
\end{figure*}

\ 
FL enables distributed devices to locally train complete models on their own data and share only the resulting model updates with a central server, instead of transmitting raw data. As depicted in Fig. \ref{fig:ArchitectureAndComparison}, central aggregation typically follows strategies like FedAvg  \cite{fedavg}, where each client's contribution is weighted by their data size. Albeit its enhancement in privacy friendliness, classical FL suffers from excessive dependence on the single central server with minimal failure tolerance. Contingent on device-to-device communication, decentralized alternatives emerge. However, both approaches struggle with computational-intensive demands for training increasingly larger models on resource-constrained nodes \cite{OpenAI2020scalinglaw} and the ``straggler problem" in compulsory aggregation synchronization, where slower nodes disrupt the entire process. 

\subsubsection{Split Learning} \
Split Learning \cite{sl} optimizes for resource-constrained scenarios by offloading most Deep Neural Network (DNN) computations to a central server, while clients process only a few early layers (called client-side model) using local data. During training, clients transmit intermediate activations and labels (referred to as ``smashed data") to the server in exchange for gradients. Random Walking Snakes (RWS) \cite{RWS} bypass server reliance by finely segmenting the model and sequentially activating a set of clients (a ``snake"), where the periodically re-shuffled ``head of snake" client, which holds the training data,  offloads model segments to other clients within a snake, as depicted in Fig. \ref{fig:ArchitectureAndComparison}. Despite the reduction in local computing and memory demands, these relay-based approaches \cite{sl,RWS} significantly increase communication overhead and add susceptibility to disruptions. Additionally, they risk overfitting and catastrophic forgetting due to data heterogeneity among clients.

\subsubsection{Combination of Federated Learning and Split Learning} \ 
Various hybrid approaches that integrate FL and Split Learning are also being explored. 
SplitFed Learning \cite{thapa2022splitfed} leverages a dual-server setup, with the main server handling server-side model computations and the federal server synchronizing client-side model updates via FedAvg, to boost communication complexity through parallel processing. 
Meanwhile, Accelerated FL \cite{han2021accelerating} reduces reliance on communication by adopting a local-loss-based training method. Notably, it utilizes two different local loss functions (one for the auxiliary network [e.g., Multi-Layer Perceptrons] connected to the cut layer on the client-side model, and the other for the server-side model's output layer) for separately updating two split models, which are further concatenated to form a final model. Therefore, it avoids receiving gradients from the main server and reduces the need for real-time synchronization. Nevertheless, the federal server-related communication problem persists. 

\vspace{-0.2cm}
\subsection{Key Issues of Distributed Learning within 6G networks}
\label{sec:Challenges}
After reviewing representative distributed learning solutions, we can highlight several underlying deployment issues, which are analyzed at the bottom part of Fig. \ref{fig:ArchitectureAndComparison} and provide the fundamental incentives for developing Snake Learning. 
\subsubsection{Reliance on Communication Synchronization}\
Wireless connections between UE and Base Stations (BSs) often fluctuate due to environmental conditions, device density, and mobility, leading to unpredictable changes in uplink and downlink speeds that disrupt the real-time data and model synchronization for distributed learning. This, in turn, exacerbates pressure on the bandwidth-limited air interface. 
Besides, asynchronous methods additionally introduce complexities like ``model staleness" due to delayed updates, which leads to model inconsistency and non-convergence, ultimately destabilizing training processes \cite{kuang2023federatedscope}. 
Hence, the over-reliance on synchronous communication heavily hinders the applicability of federal aggregation-based distributed learning frameworks. 

\subsubsection{Heterogeneous, Dynamic and Limited Resource Availability}\
Unlike stable and dedicated cloud computing resources, the inter-service sharing nature in 6G networks underscores the ``tidal effect'' of traffic loads from conventional services, implying a tidal shift in the available computational resources (e.g., CPU, memory, or any type of accelerators such as GPUs) for CaaS and AIaaS. 
Besides, the computing nodes, supplied by different vendors, are constrained by heterogeneous computational capabilities due to the limitation of hardware, processing power, storage, and energy consumption.
These result in significant disparities across computing nodes and make the aforementioned distributed learning frameworks fall short of processing computation-intensive tasks promptly and dynamically. 

\subsubsection{Heterogeneity of Data}\
Data from diverse network nodes exhibit unique distributions, commonly known as Non-Independent and Identically Distributed (Non-IID), due to differences in modalities, user behavior, and temporal \& geographic preferences. 
Such heterogeneity introduces biases during model training, affecting the stability of the training process and diminishing the generalization ability of the model \cite{abdelmoniem2023comprehensive}. 
To counteract these issues, simply adopting data-level techniques, such as augmentation or synthesizing under-represented class samples, proves inadequate. Instead, more advanced strategies must be incorporated into the design of distributed learning frameworks.

\section{Snake Learning: A Distributed Collaborative Learning Framework in 6G Networks}

This section outlines the key components and implementation workflow of Snake Learning.

\vspace{-0.3cm}
\subsection{Overview}\
Towards efficient model training/fine-tuning while addressing emerging issues in Section \ref{sec:Challenges}, Snake Learning shifts the paradigm from communication-intensive exchange of complete model updates in FL and real-time smashed data in Split Learning. 
In Snake Learning, each node with distinct local data is designated to train specific middle layers, along with the first and last layers of a DNN model, since the first layer extracts fundamental features while the last layer tailors task-specific decision boundaries for new tasks. Updated parameters are uploaded only once after completing local training on a node. 
This layer assignment for distributed training facilitates the adaptation to resource availability and heterogeneity, and locally training partial layers greatly reduces computing, storage, and communication requirements on individual nodes while maintaining privacy friendliness. 
In case of communication and computation disruptions, the chain of training can seamlessly transit to an alternative idle node, thus ensuring continuity of model training. 

\begin{figure*}[tbp]
	\centering 
	\hspace*{-0.2cm}
	\includegraphics[scale = 0.8]{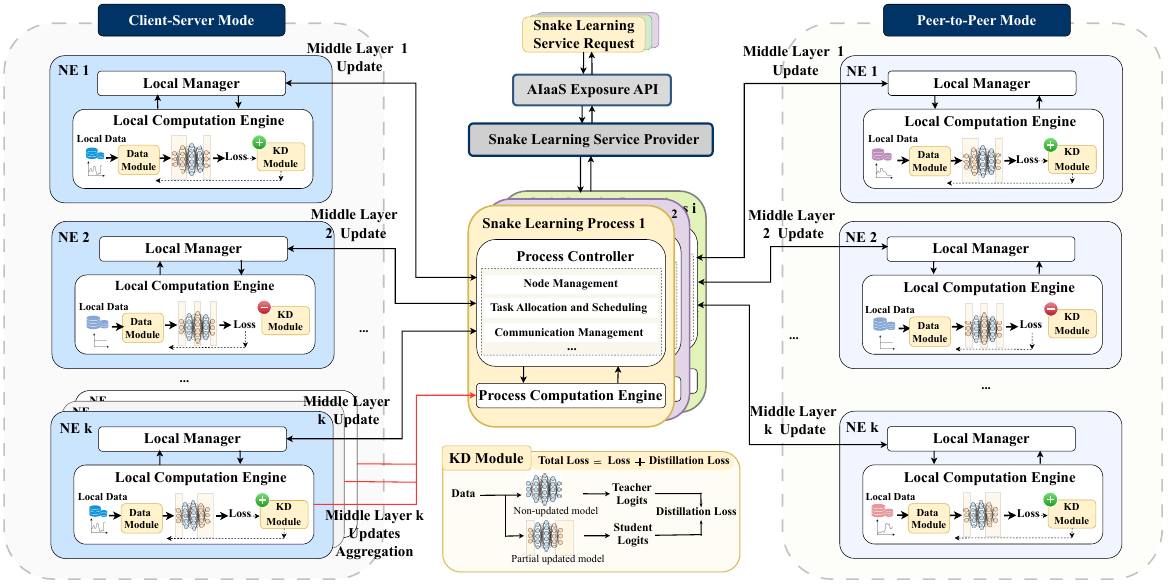}
	\caption{Workflow of Snake Learning in both Client-Server and Peer-to-Peer modes. The KD module, an abbreviation for Knowledge Distillation, activates or deactivates according to the inter-node data heterogeneity. }
 \vspace{-.2cm}
	\label{fig:operation details}
\end{figure*} 

\vspace{-0.2cm}
\subsection{{Key Enabling Components for Snake Learning}} 
\label{sec:componenents}

In line with AIaaS like FL \cite{Reviewer1SuggestPaper}, Snake Learning necessitates the integration of several enabling components within 6G networks. As the sub-architecture of distributed AI services advocated in Hexa-X project \cite{Reviewer1SuggestPaper}, these components primarily encompass the Service Provider (SP), Process Controller (PC) \& Process Computation Engine (PCE), and Local Manager (LM) \& Local Computation Engine (LCE). 

\subsubsection{Service Provider}\ 
The SP, whether located externally or within 6G networks, manages distributed training and fine-tuning services that can be instantiated on demand. 
Operators or third parties interact with the SP via the Application Programming Interface (API), which supports the registration and de-registration of Snake Learning services. 
The SP handles different user requests, creating a dedicated process for each emerging task, 
associated with essential entities (e.g., Virtual Machines [VMs] or containers) required for a library of AI models. 
At the system level, the SP orchestrates multiple concurrent processes, each managed by its own PC, allowing for simultaneous multi-task execution across the network. 
By continuously monitoring the working status of active processes, the SP can dynamically reallocate and migrate the entities that execute these processes based on service demands, resource availability, or fluctuations in wireless link quality.

\subsubsection{Process Controller \& Process Computation Engine}\ 
Each active Snake Learning process includes a dedicated PC and a PCE that can be virtually taken charge of by the SP. 
The PC bridges the SP and LMs within different computing nodes (e.g., UE), while the PCE handles computations. 
The PC is responsible for balancing tradeoffs between resource cost and performance, and is conceptionally composed of several logically separated network functions, such as node management, training task allocation \& scheduling (i.e., assigning different layers to specific nodes), and communication management.

\subsubsection{Local Manager \& Local Computation Engine}\
Each computing node is endowed with a Snake Learning LM, which manages interactions between Snake Learning services and the LCE within the node. 
To facilitate the orchestration of resources, idle UEs may request authorization from the SP in advance, while other in-network computing nodes semi-actively apply the joining. 
In terms of computational capability, resource availability, and data relevance \& quality, the PC in the process instantiated for a specific training task coordinates and determines the acceptance of nodes. Once approved, the node is added to a node management pool within the specific PC. 
Notably, during its enrollment in a specific task, each node still has the privilege to be accepted and reassigned by another PC based on its resource availability and the emerging computing demand. 
On the other hand, the LCE is responsible for performing actual computation tasks, including data processing, model training and parameter updates, and adopting appropriate measures (e.g., provisionally activating a KD module) to maintain acceptable performance.

\vspace{-0.3cm}
\subsection{Workflow of Snake Learning}
\begin{figure*}[!ht]
	\centering
 \hspace*{-0.4cm} 
	\includegraphics[scale =0.8]{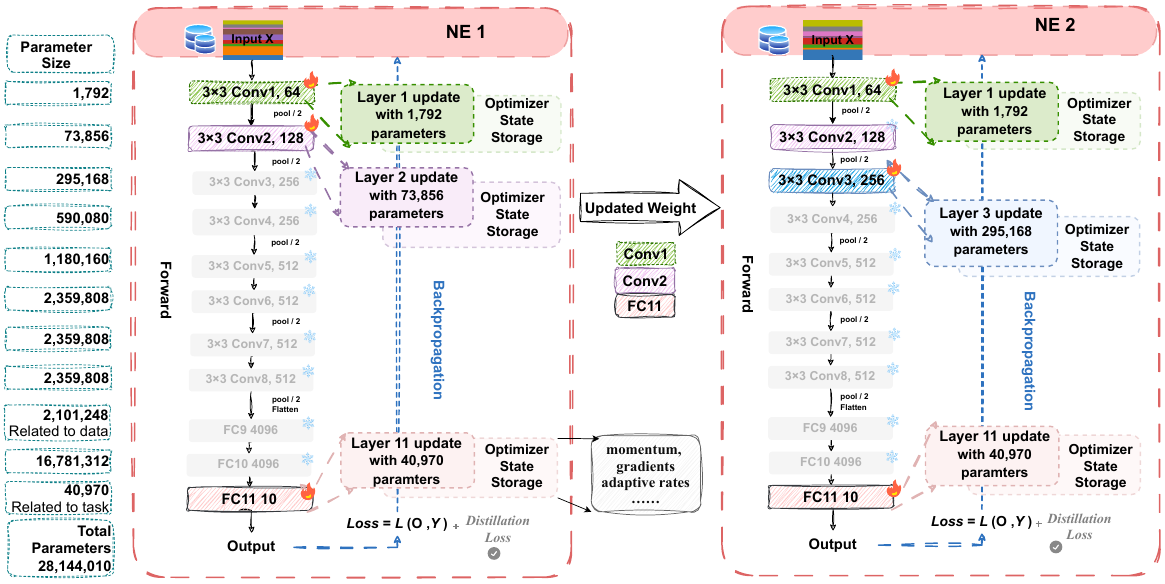}
	\caption{The illustration of Snake Learning's feasibility on VGG-11 model.}
	\label{fig:VGGAnalysis}
	\vspace{-.35cm}
\end{figure*} 

\subsubsection{Initialization}\
Snake Learning process, an instance of service registered via AIaaS API with key service descriptions such as task, objectives, requirements, and possible constraints (e.g., Service Level Agreement [SLA], data \& devices, and layer assignment guidelines), 
begins with initialization of the AI model and training tasks, and authorized participating nodes from the SP. 
Then the allocated PC for specific task takes charge of unified service node management, including identifying and verifying node connectivity, assessing and confirming node computational resources, and monitoring its status, resource usage, data quality and performance metrics. 
Additionally, to address data heterogeneity and prevent catastrophic forgetting, the KD module, which computes a distillation loss, is conditionally activated according to the measured inter-node data distribution discrepancy (e.g., Kullback–Leibler divergence, Jensen–Shannon distance or hyperparameter fitting differences).

\subsubsection{Layer Assignment}\
Once registered for a Snake Learning service, each idle node is informed by the PC of the specific ``middle layer" to be learned (i.e., one of the $2$nd to $N-1$-th layers in a model with $N$ layers). 
In other words, the PC takes charge of the layer assignment. 
Specifically, with training objective (e.g., accuracy) as the primary goal, this layer assignment, which shall account for various factors from training complexities of middle layers to network conditions (e.g., connectivity quality \& stability, and routing paths) and system constraints (e.g., bandwidth, latency, energy consumption, and resource availability), aims to make assigned layers proportionate to the computational capability and communication connectivity of nodes. Categorized as the well-studied multi-objective optimization problem, it can be solved by methods like mixed-integer linear programming, AI-driven approaches such as reinforcement learning, and bio-inspired heuristic algorithms. 
Notably, in CS mode, it is feasible to simultaneously enable multiple nodes for parallel execution of the same training task, while PCE aggregates the results. Nevertheless, this parallelization comes at the cost of compromised relaxation to synchronization requirements. 
Furthermore, fully without the involvement of the PCE, the P2P mode might also involve the slight participation of the PC. For example, the first training node acquires an initial model and tasks from the PC, while next-hop nodes are coordinately selected by the PC.

\subsubsection{Local Training}\
Same as FL, Snake Learning adopts privacy-friendly distributed data training by keeping raw data at generated nodes only. Additionally, advanced techniques like differential privacy and homomorphic encryption can be further employed to strengthen the privacy-preserving capabilities of Snake Learning by reducing data leakage risks and defending against malicious attacks during parameter transmission. 
Before training the assigned layers from locally stored data, a Data Module in LCE gets involved for preprocessing (e.g., normalization, cleaning) and filtering of the data. 
To ensure training effectiveness, local data heterogeneity can be optimized by adopting techniques such as clustering and anomaly detection to identify similarities and isolate data inconsistencies, while inter-node heterogeneity that leads to excessive imbalances in updates can be alleviated by KD module. 
In that regard, if the KD module is triggered in initialization, it conducts an extra distillation operation by calculating the cross-entropy loss. This loss measures the discrepancy between the model previously received by the node (as the ``teacher") and the model currently undergoing local training iterations (as the ``student"), thus preventing catastrophic forgetting while maintaining learning stability. 
Besides, integrating gradient clipping and adaptive learning rate with optimizers like Adam, RMSprop can contribute to improving learning effectiveness. In our following serpentine training feasibility study, the initial learning rate for updating the parameters of each middle layer remains constant, while the adjustment (e.g., decaying the learning rate) is performed after updating all layers once (i.e., one cycle), thus leading to uniform inter-layer learning speed and boosting model stability across cycles. 
Meanwhile, prior to updating the model, the computed gradients undergo a clipping operation to avoid gradient explosion. 
In addition, those non-updated layers' parameters can be quantized to further reduce storage demands. 

\vspace{-0.1cm}
\subsubsection{Updated Parameter Transmission and Node Management}\
Normally, in CS mode, the updated parameters are uploaded to the PCE, which aggregates (if necessary) and disseminates them to the next node selected by the PC. 
In P2P mode, nodes also learn the next hop node from the PC, and if the next node has the latest model cached, updated parameters can be directly transmitted. 
For each computing node, communication for parameter uploading occurs only once at the end of local training. 
Nodes can proactively request the PC to exit a Snake Learning service upon reaching the designated training epochs, predefined threshold training metrics, or detecting trivial performance improvement. 
The task-specific PC continuously monitors each accepted node's status and retains the authority to remove nodes from the training process and inform SP under conditions such as intermittent and unstable network connectivity, unexpected deterioration in resource availability (e.g., load, memory, battery status), or disqualified local training efficiency (e.g., unsatisfactory processing speed or suboptimal training accuracy). 
When this occurs, the PC ensures continuity by notifying the node's LM to terminate its participation and initiate updated parameter upload, while dynamically reallocating the training task to candidate nodes with available resources. 
This process helps maintain training consistency and robustness, ensuring Snake Learning remains scalable and adaptable to the dynamic complexities of 6G networks. 

\vspace{-0.2em}
\section{Feasibility Study of Snake Learning}
This section presents the feasibility study of Snake Learning's serpentine training and highlights its superiority over existing representative methods on classic tasks including the widely-used image classification task in the literature as well as emerging more complex use cases like LLM fine-tuning. 

\vspace{-.3cm}
\subsection{Feasibility Study}
\label{sec:Feasibility}
\subsubsection{Training of Classification Tasks}\label{sec:VGG}
\begin{figure}[!tbp]
	\centering 
	\hspace*{-0.2cm}
	\includegraphics[scale = 0.3]{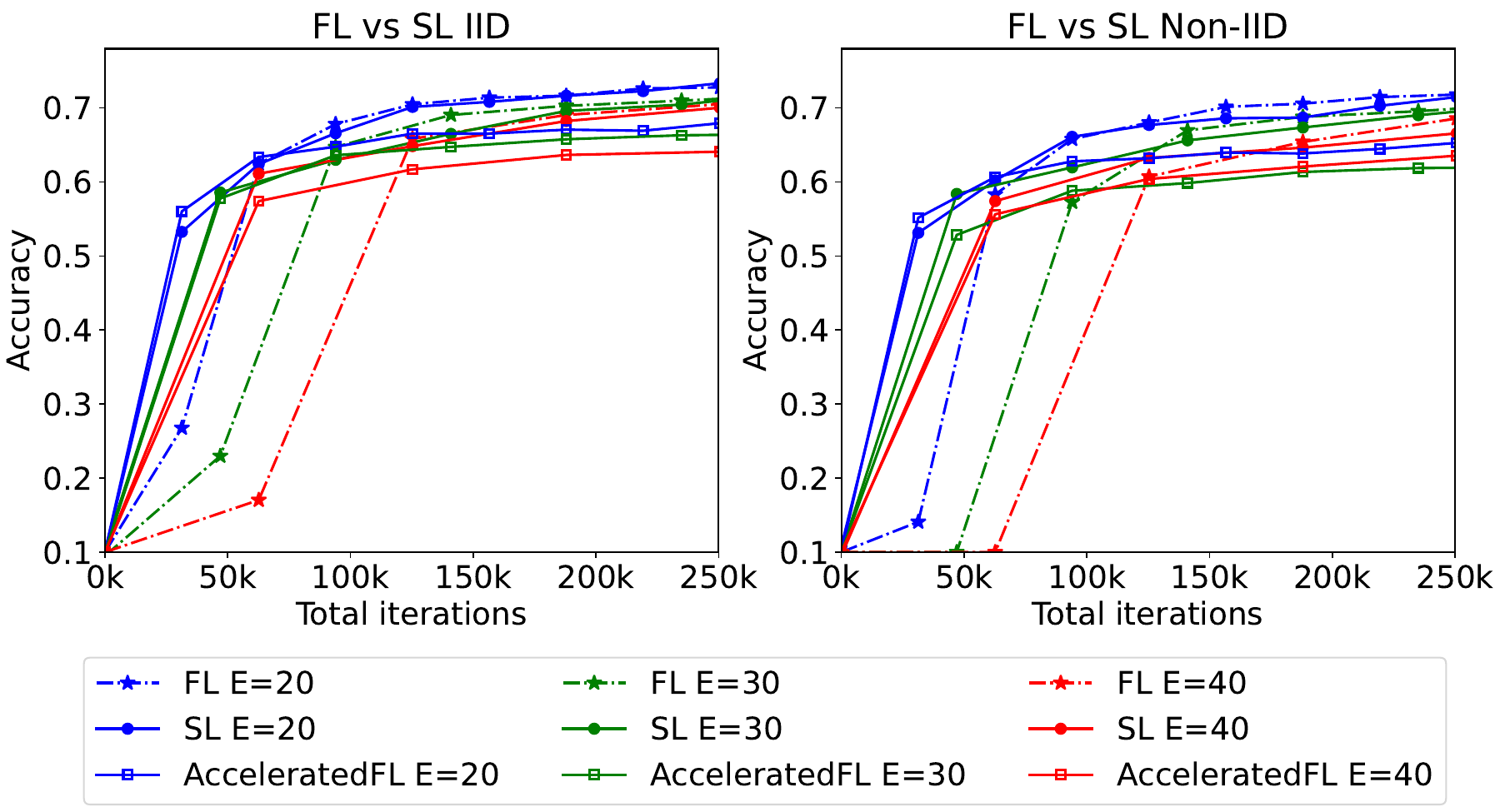}
	\caption{Image classification performance comparison of other frameworks and Snake Learning (SL) across varying training epochs $E$. 
  }
	\label{fig:CVresults}
	\vspace{-.35cm}
\end{figure}

\
We demonstrate the effectiveness of Snake Learning using the VGG-11 model on the CIFAR-10 dataset. 
VGG-11 comprises $8$ Convolutional (Conv) and $3$ Fully Connected (FC) layers, with the last FC layer producing a $10$-class output. 
As illustrated in Fig. \ref{fig:VGGAnalysis}, Snake Learning assigns the $9$ middle layers (excluding the $1$st and $11$th) to distributed nodes with distinct local data, whereas Accelerated FL updates the first two layers of model and an auxiliary linear layer and FL updates the entire model across the same nodes.  
In IID cases, $50,000$ samples (each with $32\times 32$ pixels) are uniformly spread among $9$ nodes. 
For Non-IID scenarios, the datasets are partitioned based on Dirichlet distribution as class priors. We allocate data $D_k$ to $k$-th node based on a sampled $D \sim \text{Dir}(\alpha)$, where $\alpha$ determines the degree of Non-IID, with a default value of $2.0$. 
One epoch (i.e., $E=1$) refers to one entire passing of a node's local dataset. 
After the designated training epochs, 
Accelearted FL and FL executes FedAvg aggregation \cite{fedavg}, while Snake Learning transfers the updated parameters. 
Results in Fig. \ref{fig:CVresults} show that Snake Learning rapidly attains $60\%$ accuracy with significantly fewer training iterations, underscoring its computational efficiency for individual nodes. 
More importantly, Snake Learning achieves acceptable final performance (i.e., exceeding $95\%$ of that of FL), while simultaneously reducing communication overhead per communication round by nearly half, as detailed in Fig. \ref{fig:ArchitectureAndComparison}. Furthermore, our experiment experience also indicates that Snake Learning provides appealing performance robustness in dynamic wireless networks and the unexpected interruption during layer-wise communications only leads to trivial performance degradation.
These results corroborate Snake Learning's promise for distributed serpentine learning with limited communication and computational resources only. 

Additionally, we further examine layer update sequences, finding that reverse middle layer training (i.e., from deep to early layers) violates hierarchical feature learning. Random and sequential updates yield similar results in IID cases, while sequential serpentine training is superior in Non-IID settings. 
Such observations align with common DNN insights that early layers learn basic features and deeper layers capture complex patterns, guaranteeing Snake Learning's robustness to heterogeneous data distributions.

\subsubsection{Fine-Tuning of LLMs}\
\label{sec:fine-tuning of LLMs}
Distributed fine-tuning of data-hungry and compute-intensive LLMs within 6G network nodes becomes crucial, yet challenging due to the single node's memory usage limitation in FL and billions of communication demands in Split Learning. 
Thus, such LLMs can gain from Snake Learning by optimizing the fine-tuning process across multiple nodes with lower computing, memory, and communication demands. 
Our experiments employ Supervised Fine-Tuning (SFT) on both the $1.3$-billion-parameter OPT model and the $8$-billion-parameter Llama-3 model, featuring $24$ and $32$ transformer blocks respectively, for causal language tasks. 
Apart from the embedding layer and the linear output layer for generating predictions, the intermediate transformer blocks are assigned to distinct nodes for fine-tuning, with each node updating a specific transformer block and employing strategies like freezing or $8$-bit quantization on non-updated parameters to further reduce memory. 
Besides, rather than simply using the typical cosine learning schedule strategy in LLMs, Snake Learning maintains a constant learning rate throughout the cycle updating of all blocks, while using the systematic decay for the next cycle to guarantee consistent training across layers. Experimental results show that such a design contributes to avoiding overfitting and enhancing model stability. 
To further enhance computing and memory efficiency, we complement Snake Learning with the parameter-efficient Low-Rank Adaptation (LoRA) technique \cite{NetGPT}, decomposing weight updates for all linear layers into the product of two low-rank matrices, thus significantly reducing the complexity of the model while maintaining performance. 
Additionally, given the variability in computing power among distributed nodes, assigning a personalized rank to each node could enable a more efficient balance between model size, computational cost, and performance. This aspect warrants further exploration. 
\begin{figure}[!tbp]
	\subfigcapskip = -1pt 
	\begin{center}
		\hspace*{-0.3cm} 
		\includegraphics[scale = 0.3]{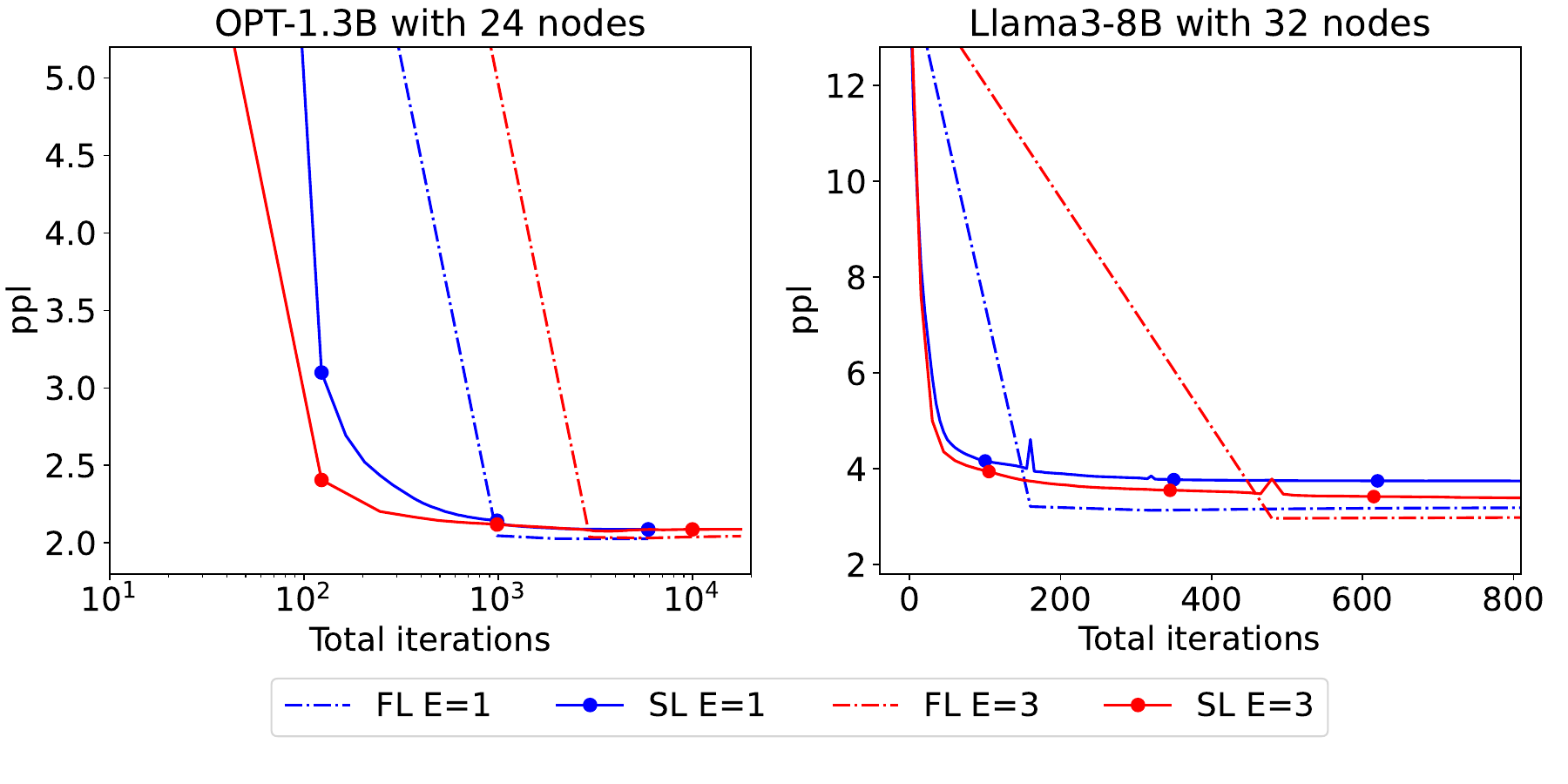}
		\caption{LLM fine-tuning performance comparison of FL and Snake Learning (SL) across varying training epochs $E$, where the performance is measured by ppl with lower values indicating greater model certainty and effectiveness. } 
		\vspace{-2em}
		\label{fig:results-of-llm}
	\end{center}
\end{figure} 
$4$ query-answer pair datasets including rm-static, synthtic-instruct-gptj-pairwise, full-hh-rlhf, and rlhf-reward-datasets are used for OPT-1.3B, while higher-quality dataset ``ChatGPTGroundTruth" consisting of $160$ sub-topics are used for Llama3-8B. 
The datasets are uniformly distributed across the nodes in IID settings; while in Non-IID settings, each node holds different datasets or different types of sub-topics. 
Fig. \ref{fig:results-of-llm} provides the comparison between Snake Learning and FL, showing that compared to FL, Snake Learning achieves a more rapid decline in perplexity (ppl) in IID cases. Besides, as validated by experiments, the performance superiority holds for Non-IID settings as well, indicating its efficient utilization of nodes' data to enhance model performance. 
Furthermore, this performance improvement becomes increasingly apparent as the number of epochs increases. 
Meanwhile, the memory footprint for fine-tuning the OPT-1.3B model on a single node is reduced from approximately $19.37$ GB in conventional FL to just $3.13$ GB for Snake Learning. 
These results indicate that Snake Learning is a promising framework for collaboratively fine-tuning LLMs on resource-constrained network nodes.

\vspace{-0.3cm}
\subsection{Discussions}
Besides the performance improvement shown in Section \ref{sec:Feasibility}, it is worthy to discuss the following aspects of Snake Learning.

\subsubsection{Relaxed Synchronization Requirements}\
Snake Learning employs sequential learning, performing complete iterations on individual nodes without the need for synchronization, as required in FL for parameter aggregation or Split Learning for intermediate activation/gradient transmission. This enables flexible training schedules such as off-peak time fine-tuning. 

\subsubsection{Computation Savings}\
While non-updated layers require forward propagation to pass features, their weight gradient computations can be skipped, retaining only activation gradients for backpropagation. This reduces FLOPs and leads to computational savings on-node. 
Integrating pruning and quantization can further enhance these savings.

\subsubsection{Memory Savings}\ 
Snake Learning significantly lowers memory usage by storing only updated parameter gradients and their optimizer states, while these factors deeply affect the peak memory occupation during training. 
Quantization of non-updated parameters can lead to additional savings. 

\subsubsection{Communication Savings}\
Split Learning's communication overhead is proportional to the data size, resulting in large volumes of smashed data, especially for fine-tuning LLMs. 
Meanwhile, due to the real-time synchronization of the complete model, significant communication overhead is also required in FL. 
In comparison, Snake Learning transfers only locally updated partial parameters and eliminates the need for frequent synchronization, thus saving significant communication overhead, especially as local iterations increase.

\subsubsection{Data Heterogeneity Adaptation and Scalability} 
As shown in Fig. \ref{fig:CVresults} and Fig. \ref{fig:results-of-llm}, Snake Learning excels with Non-IID data through techniques, such as KD, gradient clipping \& learning rate adjustment, and better aligns with the real-world scenarios of gradually garnering data and training the model.

\subsubsection{Training Time} 
The sequential training nature in Snake Learning takes increased training time. Therefore, it is more suitable for some time-insensitive distributed training/fine-tuning tasks, to reap the available computing resources for off-peak communication services. 

\subsubsection{Interoperability} 
As discussed in Section \ref{sec:componenents}, APIs are essential for AIaaS due to the complication of orchestrating distributed communication and computing resources, while reserved APIs need further investigation to implement task-oriented orchestration and guarantee service continuity. 
Initiatives such as the ONAP MultiCloud project and Camara's Edge Cloud, which enables the exposure of resources and features for optimizing VNF (Virtual Network Function) homing and placement and the deployment of applications on VMs and containers, are exemplary in this regard. 

\vspace{-0.2em}
\section{Conclusions \& Open Research Directions}\label{sec:conclusion}

We propose the distributed learning framework ``Snake Learning" that optimizes computational resource utilization within 6G by assigning partial layers for sequential model training/fine-tuning across idle nodes. 
Its serpentine training mechanism minimizes synchronization and data transmission, addressing the limitations of traditional FL and Split Learning. Knowledge distillation and gradient clipping help prevent catastrophic forgetting caused by data heterogeneity and gradient explosion. 
Evaluations on classification and LLM fine-tuning tasks confirm that it outperforms in decentralizing workloads and reducing storage, memory, and communication demands on individual nodes, while maintaining robust model performance. While Snake Learning is a promising distributed learning framework for provisioning AI model training/fine-tuning services, there are still some remaining open issues, such as open API design and standardization for enhanced interoperability, fine-grained layer assignment schemes for a library of typical AI models, thorough node selection and resource scheduling management schemes. 

\vspace{-0.5em}
\bibliographystyle{IEEEtran}
\bibliography{reference}

\section*{Author Biographies}
\textbf{Xiaoxue Yu} (sdwhyxx@zju.edu.cn)  is a PhD Candidate in Zhejiang University, Hangzhou, China. Her research interests currently focus on communications in distributed learning. 

\textbf{Xingfu Yi} (yixingfu@zju.edu.cn) was a master student in Zhejiang University, and graduated in Mar. 2024. 

\textbf{Rongpeng Li} (lirongpeng@zju.edu.cn) is an Associate Professor in Zhejiang University. His research interests currently focus on networked intelligence. 

\textbf{Fei Wang} (wangfei76@huawei.com) is a Chief Researcher of Huawei Technologies. His research directions include 6G wireless network architecture and distributed learning, etc.

\textbf{Chenghui Peng} (pengchenghui@huawei.com) is a Principal Researcher of Huawei Technologies. His current research interests focus on 6G native AI architecture design.

\textbf{Zhifeng Zhao} (zhaozf@zhejianglab.com) is the Chief Engineer with Zhejiang Lab, Hangzhou, China. His research area includes collective intelligence and software-defined networks.

\textbf{Honggang Zhang} (hgzhang@cityu.edu.mo) is a Professor in City University of Macau, Macau, China. He is interested in cognitive green communications.

\section*{Supplementary Results}

\begin{figure}[h]
    \centering 
    \begin{minipage}[t]{0.48\textwidth}
        \centering
        \hspace*{0.4cm} 
        \includegraphics[scale=0.26]{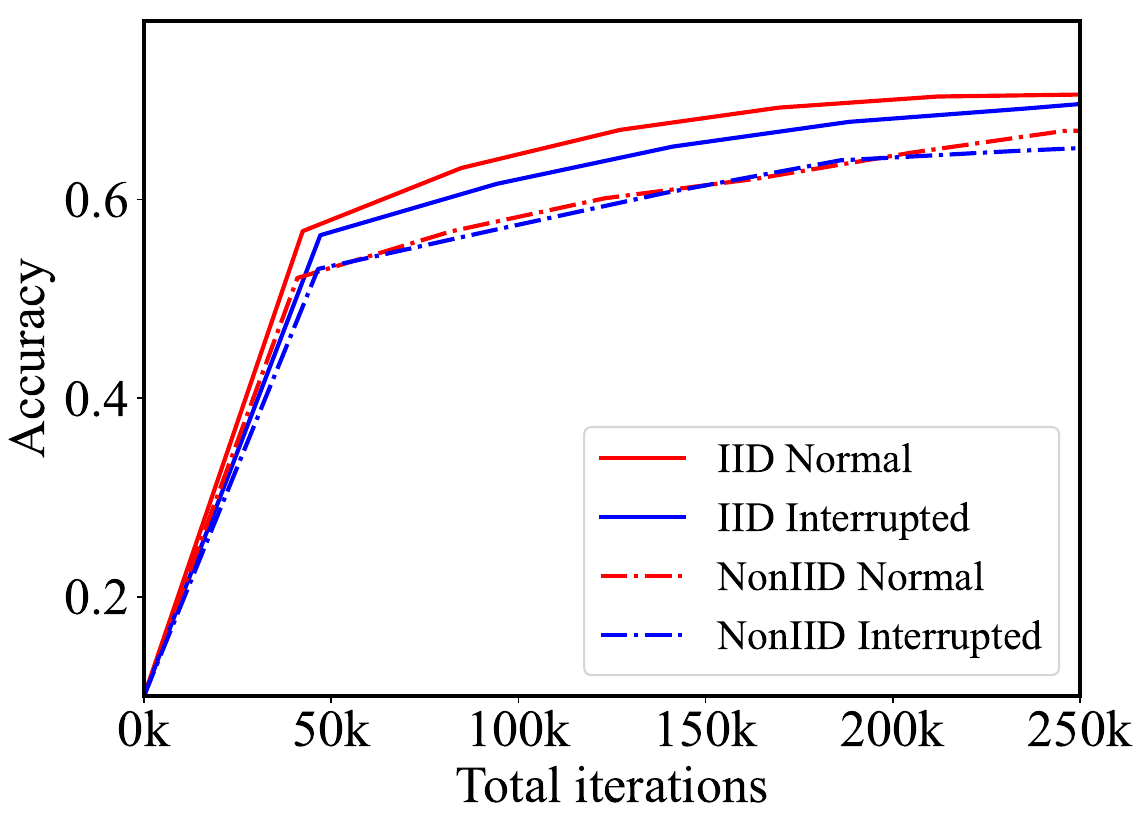}
        \caption{Performance under the effect of training interruption.}
        \label{fig:A1_interruption}
    \end{minipage}
    \hfill
    \vspace{1.8em}
    \begin{minipage}[t]
    {0.48\textwidth}
        \centering
        \hspace{
        -1.5em
        }
        \includegraphics[scale=0.48]{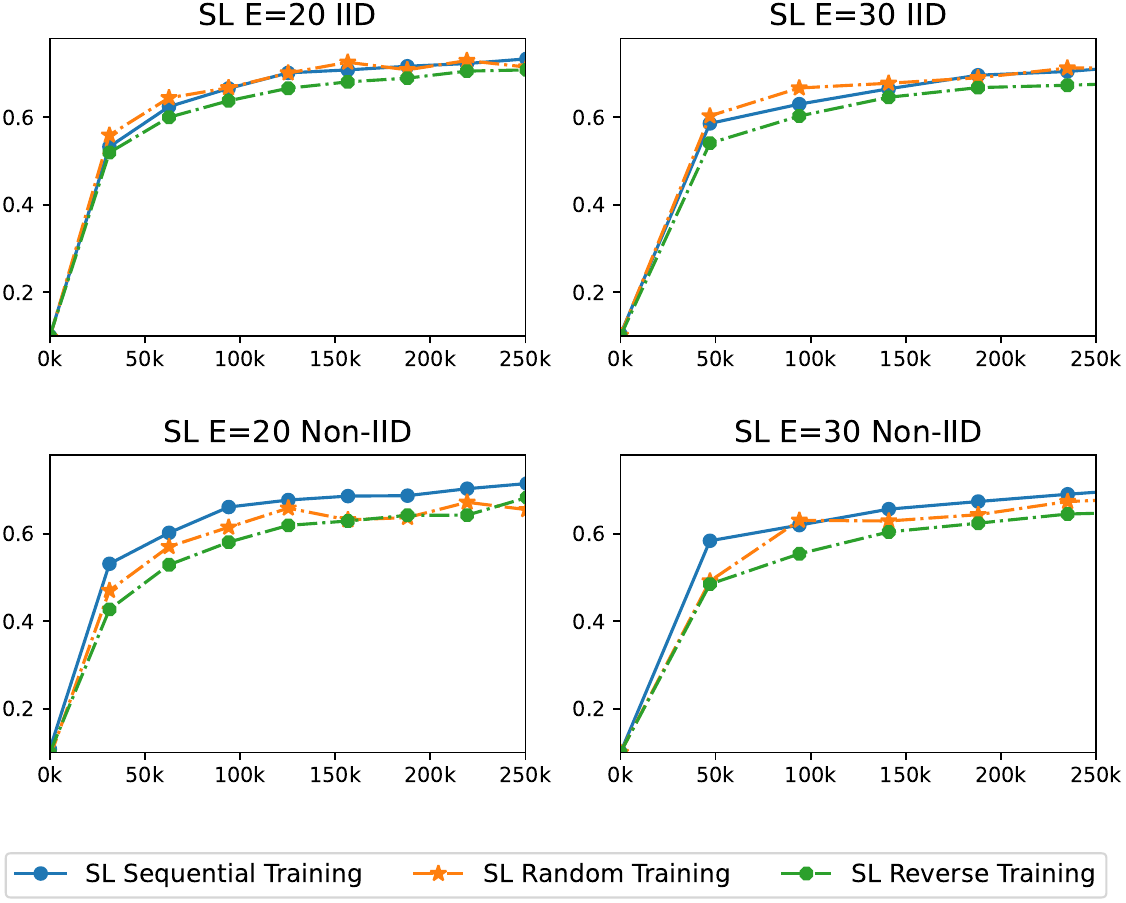}
        \caption{Performance comparison of different training sequences.}
        \label{fig:A1_sequence}
    \end{minipage}
\end{figure}

Fig. \ref{fig:A1_interruption} highlights the impact of interruptions and the seamless transition of the training chain to an idle node, as discussed in Section \ref{sec:VGG}. 
Snake Learning demonstrates robust performance in dynamic wireless networks, with interrupted training on a specific layer (5th layer of VGG-11 in this experiment), which is labeled as ``Interrupted", achieving $98.8\%$ and $97.8\%$ of the performance obtained by the uninterrupted ``Normal" serpentine training across different nodes under IID and Non-IID data distributions, respectively. 
Despite minor degradation caused by layer-wise communication interruptions, the overall performance remains stable and resilient. 


Fig. \ref{fig:A1_sequence} presents a detailed comparison of various layer-wise training sequences as discussed in Section \ref{sec:VGG}. It is evident that the sequential/serpentine training strategy of Snake Learning guarantees consistent performance and robustness across both IID and Non-IID data distributions.

\begin{figure}[t]
\subfigcapskip = -1pt
\begin{flushleft}
\noindent
\includegraphics[scale=0.3]{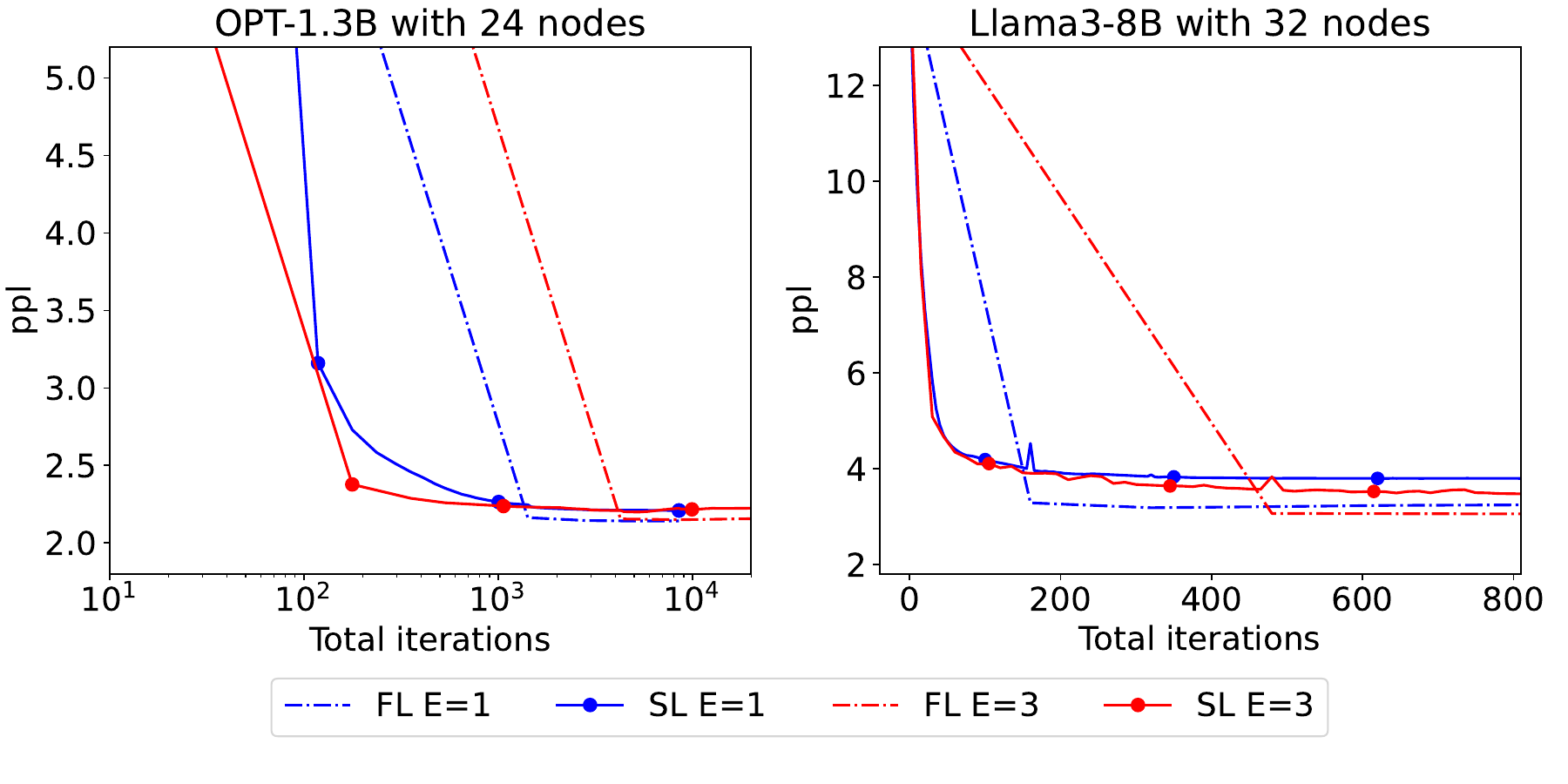}
\caption{LLM fine-tuning performance comparison of FL and Snake Learning (SL) across varying training epochs $E$ under Non-IID settings.}
\vspace{-1.5em}
\label{fig:results-of-llm-noniid}
\end{flushleft}
\end{figure}

In Fig. \ref{fig:results-of-llm-noniid}, the comparison between FL and Snake Learning (SL) under Non-IID settings in LLM fine-tuning tasks, as discussed in Section \ref{sec:fine-tuning of LLMs}, clearly demonstrates SL's advantage in reducing ppl more rapidly, underscoring its ability to effectively utilize nodes’ data in Non-IID cases and enhance model performance. 

\end{document}